\documentclass[11pt,a4paper]{article}

\usepackage[utf8]{inputenc}
\usepackage[T1]{fontenc}
\usepackage{lmodern}
\usepackage[margin=1in]{geometry}
\usepackage{hyperref}
\usepackage{graphicx}
\usepackage{listings}
\usepackage{xcolor}
\usepackage{booktabs}
\usepackage{amsmath}
\usepackage{url}
\usepackage{natbib}
\usepackage{tikz}
\usetikzlibrary{arrows.meta, positioning}

\definecolor{codebg}{RGB}{248,248,248}
\definecolor{codeframe}{RGB}{200,200,200}
\definecolor{commentcolor}{RGB}{100,100,100}
\definecolor{keywordcolor}{RGB}{0,0,180}
\definecolor{stringcolor}{RGB}{0,128,0}

\lstdefinelanguage{DriftScript}{
  morekeywords={believe,ask,goal,inherit,similar,imply,predict,equiv,instance,
    seq,and,or,not,product,call,cycles,def-op,reset,config,concurrent,
    ext-set,int-set,ext-inter,int-inter,ext-diff,int-diff,
    ext-image1,ext-image2,int-image1,int-image2},
  sensitive=true,
  morecomment=[l]{;},
  morestring=[b]",
  alsoletter={-},
  literate={:now}{{\textcolor{keywordcolor}{:now}}}3
           {:past}{{\textcolor{keywordcolor}{:past}}}4
           {:future}{{\textcolor{keywordcolor}{:future}}}6
           {:truth}{{\textcolor{keywordcolor}{:truth}}}5
           {:dt}{{\textcolor{keywordcolor}{:dt}}}2,
}

\lstdefinelanguage{EBNF}{
  morekeywords={},
  sensitive=true,
  morecomment=[l]{},
  morestring=[b]",
  literate={|}{{\textcolor{keywordcolor}{|}}}1
           {=}{{\textcolor{keywordcolor}{=}}}1
           {;}{{\textcolor{keywordcolor}{;}}}1,
}

\title{DriftScript: A Domain-Specific Language for Programming\\Non-Axiomatic Reasoning Agents}

\author{
  Seamus Brady\\
  Independent Researcher\\
  Dublin, Ireland\\
  \texttt{seamus@corvideon.ie}\\
  \url{https://seamusbrady.ie}
}

\date{}

\begin{document}

\maketitle

\begin{abstract}
Non-Axiomatic Reasoning Systems (NARS) provide a framework for building
adaptive agents that operate under insufficient knowledge and resources.
However, the standard input language, Narsese, poses a usability barrier: its
dense symbolic notation, overloaded punctuation, and implicit conventions make
programs difficult to read, write, and maintain. We present DriftScript, a
Lisp-like domain-specific language that compiles to Narsese. DriftScript
provides source-level constructs covering the major sentence and term forms
used in Non-Axiomatic Logic (NAL) levels 1 through~8---including inheritance,
temporal implication, variable quantification, sequential conjunction, and
operation invocation---while replacing symbolic syntax with readable
keyword-based S-expressions. The compiler is a zero-dependency, four-stage
pipeline in 1{,}941 lines of C99. When used with the DriftNARS engine,
DriftScript programs connect to external systems through four structured
callback types and an HTTP operation registry, enabling a sense-reason-act
loop for autonomous agents. We describe the language design and formal grammar,
detail the compiler architecture, and evaluate the compiler through a 106-case
test suite, equivalence testing against hand-written Narsese, a NAL coverage
analysis, structural readability metrics, and compilation benchmarks. The
source code is available at
\url{https://github.com/seamus-brady/DriftNARS}.

This paper focuses on the design and implementation of the DriftScript
language and its embedding into DriftNARS, rather than on new inference
algorithms for NARS itself.
\end{abstract}

\section{Introduction}

Non-Axiomatic Reasoning Systems (NARS) \citep{wang2006rigid, wang2013non}
implement a logic designed for intelligent systems that must operate with
insufficient knowledge and resources. Unlike classical logic, which assumes a
closed and consistent knowledge base, Non-Axiomatic Logic (NAL) treats every
belief as uncertain, every inference as defeasible, and every conclusion as
subject to revision in light of new evidence. These properties make NARS
attractive for building autonomous agents that must adapt to unforeseen
environments.

The standard input language for NARS is Narsese, a formal notation in which
statements combine terms through copulas (relationship operators) and receive
truth values expressing the system's degree of belief. Narsese is expressive
enough to encode inheritance hierarchies, temporal cause-effect rules,
multi-step plans, and goal-directed behaviour. However, its syntax is
difficult for newcomers and cumbersome for nontrivial programs:

\begin{lstlisting}[language={}]
<(light_on &/ <(*, {SELF}, switch) --> ^press>) =/> light_off>.
\end{lstlisting}

This single Narsese sentence encodes ``if the light is on and the agent
presses the switch, the light turns off.'' The overloaded angle brackets,
symbolic copulas (\texttt{=/>}), set notation (\texttt{\{SELF\}}), product
terms (\texttt{*}), operation prefixes (\texttt{\^{}}), and tense markers
create a high entry cost for new users. Difficulty reading and writing
Narsese discourages experimentation, slows debugging, and limits NARS
adoption outside the community of specialists who have internalised its
conventions.

Prior NARS implementations---notably OpenNARS \citep{wang1995non}
and OpenNARS for Applications (ONA) \citep{hammer2020opennars}---have focused
on the reasoning engine and its algorithmic properties, leaving the input
language unchanged from its original formal specification. We are not aware
of prior published work on a compiled DSL covering this range of Narsese
constructs.

We present \textbf{DriftScript}, a domain-specific language (DSL) that
compiles to Narsese. DriftScript makes the following contributions:

\begin{enumerate}
\item \textbf{A readable surface syntax for a broad subset of NAL~1--8.}
  DriftScript replaces Narsese's symbolic notation with Lisp-like
  S-expressions using human-readable keywords. The same temporal rule above
  becomes:

\begin{lstlisting}[language=DriftScript]
(believe (predict (seq "light_on"
                       (call ^press (ext-set "SELF") "switch"))
                  "light_off"))
\end{lstlisting}

\item \textbf{A lightweight, zero-dependency compiler.} The DriftScript
  compiler is a pure text transformer implemented as a four-stage pipeline
  (tokenise, parse, compile, emit) in 1{,}941 lines of C99. It produces
  Narsese output and engine control directives with static validation and
  source-location diagnostics.

\item \textbf{Practical embedding in a callback-driven agent runtime.} When
  used with the DriftNARS engine \citep{brady2024driftnars}, compiled scripts
  connect to external systems through C, Python, and HTTP interfaces, enabling
  agents that learn from observation, make decisions, and act on their
  environment.
\end{enumerate}

The remainder of this paper is organised as follows. Section~\ref{sec:bg}
provides background. Section~\ref{sec:lang} describes the language design,
including a formal grammar. Section~\ref{sec:compiler} details the compiler.
Section~\ref{sec:integration} presents integration mechanisms.
Section~\ref{sec:eval} evaluates the compiler.
Section~\ref{sec:examples} demonstrates agent programming through worked
examples. Section~\ref{sec:related} discusses related work.
Section~\ref{sec:conclusion} concludes.

\section{Background}
\label{sec:bg}

\subsection{Non-Axiomatic Reasoning}

Non-Axiomatic Logic (NAL) \citep{wang2006rigid} is a term logic designed for
systems that reason under the Assumption of Insufficient Knowledge and
Resources (AIKR). In contrast to classical logics that assume completeness and
consistency, NAL treats every piece of knowledge as tentative and every
inference as revisable.

NAL is organised into nine levels of increasing capability \citep{wang2013non}:

\begin{itemize}
\item \textbf{NAL-1--2}: Inheritance and similarity reasoning with
  truth-value functions.
\item \textbf{NAL-3--4}: Set operations, intersections, and products for
  structured knowledge.
\item \textbf{NAL-5--6}: Implication, equivalence, variable quantification,
  and higher-order statements.
\item \textbf{NAL-7--8}: Temporal reasoning, sequential conjunction,
  procedural knowledge, and goal-directed decision making.
\item \textbf{NAL-9}: Self-monitoring and introspection (not addressed in
  this paper).
\end{itemize}

\subsection{Narsese}

Narsese is the formal input language of NARS. Every input is a \emph{sentence}:
a term combined with a punctuation mark and an optional truth value.
Table~\ref{tab:copulas} and Table~\ref{tab:connectors} summarise the copulas
and connectors.

\begin{table}[h]
\centering
\caption{Narsese copulas}
\label{tab:copulas}
\begin{tabular}{lll}
\toprule
\textbf{Copula} & \textbf{Notation} & \textbf{Meaning} \\
\midrule
Inheritance & \texttt{-->} & ``is a kind of'' \\
Similarity & \texttt{<->} & ``resembles'' \\
Implication & \texttt{==>} & ``if \ldots\ then'' (eternal) \\
Temporal implication & \texttt{=/>} & ``if \ldots\ then'' (temporal) \\
Equivalence & \texttt{<=>} & ``if and only if'' \\
\bottomrule
\end{tabular}
\end{table}

\begin{table}[h]
\centering
\caption{Narsese connectors}
\label{tab:connectors}
\begin{tabular}{lll}
\toprule
\textbf{Connector} & \textbf{Notation} & \textbf{Meaning} \\
\midrule
Sequential conjunction & \texttt{\&/} & Events in temporal order \\
Conjunction & \texttt{\&\&} & Both hold \\
Disjunction & \texttt{||} & At least one holds \\
Negation & \texttt{-{}-} & Negation \\
Product & \texttt{*} & Ordered tuple \\
Extensional set & \texttt{\{\ldots\}} & Specific individuals \\
Intensional set & \texttt{[\ldots]} & Properties \\
\bottomrule
\end{tabular}
\end{table}

\textbf{Punctuation} determines the sentence type: `\texttt{.}' for beliefs,
`\texttt{?}' for questions, and `\texttt{!}' for goals.
\textbf{Truth values} are pairs $\{f, c\}$ where frequency~$f \in [0,1]$
indicates how often the statement holds and confidence~$c \in [0,1]$ indicates
evidence strength. The default is $\{1.0, 0.9\}$. An \emph{expectation}
value, used in DriftNARS (following ONA) for decision making, is computed as:
\begin{equation}
e = c(f - 0.5) + 0.5
\end{equation}
The engine executes an operation when expectation exceeds a configurable
threshold.
\textbf{Tense markers} stamp sentences with temporal information:
\texttt{:|:}~(present), \texttt{:\textbackslash:}~(past),
\texttt{:/:}~(future, questions only).

\subsection{OpenNARS for Applications}

ONA \citep{hammer2020opennars} is a C-based NARS implementation emphasising
real-time performance and procedural reasoning (NAL-7/8). DriftNARS
\citep{brady2024driftnars} is a fork of ONA restructured as an embeddable
library with instance-based state management and a public C~API. DriftScript
was developed as part of DriftNARS but produces standard Narsese output and
could in principle target any compatible NARS implementation.

\section{The DriftScript Language}
\label{sec:lang}

\subsection{Design Goals}

DriftScript was designed with four goals: (1)~\textbf{Readability}---replace
symbolic notation with English keywords; (2)~\textbf{Broad coverage}---provide
constructs for the major Narsese forms used in NAL~1--8;
(3)~\textbf{Static validation}---catch structural errors at compile time;
(4)~\textbf{Composability}---produce standard Narsese output that can be
mixed with raw Narsese input.

The choice of S-expression syntax was deliberate: prefix notation allows
trivial parsing with recursive descent, uniform arity checking on every form,
no ambiguity with Narsese's own punctuation characters, and a natural path to
future macro support.

\subsection{Formal Grammar}

The following grammar defines DriftScript in EBNF notation. Arity constraints
not expressible in the grammar (e.g., copulas require exactly 2 arguments,
\texttt{not} requires 1, \texttt{seq} requires 2--3) are enforced by the
compiler.

\begin{lstlisting}[language=EBNF,basicstyle=\footnotesize\ttfamily]
program     = { form } ;
form        = sentence | meta ;
sentence    = "(" sent_kw term { option } ")" ;
sent_kw     = "believe" | "ask" | "goal" ;
option      = tense | truth | dt ;
tense       = ":now" | ":past" | ":future" ;
truth       = ":truth" float float ;
dt          = ":dt" integer ;
term        = atom | copula_form | conn_form | call_form ;
atom        = string | variable | operation ;
string      = '"' { char } '"' ;
variable    = ( "$" | "#" | "?" ) ident ;
operation   = "^" ident ;
copula_form = "(" copula term term ")" ;
copula      = "inherit" | "similar" | "imply" | "predict"
            | "equiv" | "instance" ;
conn_form   = "(" connector term { term } ")" ;
connector   = "seq" | "and" | "or" | "not" | "product"
            | "ext-set" | "int-set" | "ext-inter" | "int-inter"
            | "ext-diff" | "int-diff" | "ext-image1" | "ext-image2"
            | "int-image1" | "int-image2" ;
call_form   = "(" "call" operation { term } ")" ;
meta        = "(" "cycles" integer ")" | "(" "def-op" operation ")"
            | "(" "reset" ")" | "(" "config" config_key value ")"
            | "(" "concurrent" ")" ;
comment     = ";" { any } newline ;
\end{lstlisting}

\subsection{Sentence Forms}

The three sentence forms correspond to Narsese sentence types:

\begin{lstlisting}[language=DriftScript]
(believe <term>)                  ; belief  -> "."
(believe <term> :now)             ; present-tense -> ". :|:"
(believe <term> :truth 0.8 0.9)  ; truth value -> ". {0.8 0.9}"
(ask <term>)                      ; question -> "?"
(goal <term>)                     ; goal (always present) -> "! :|:"
\end{lstlisting}

Options can be combined. Goals are always emitted with present tense;
questions may use any tense.

\subsection{Terms and Copulas}

Copulas are binary prefix operators. Table~\ref{tab:ds_copulas} shows the
mapping.

\begin{table}[h]
\centering
\caption{DriftScript copulas and their Narsese equivalents}
\label{tab:ds_copulas}
\begin{tabular}{lll}
\toprule
\textbf{DriftScript} & \textbf{Narsese} & \textbf{Meaning} \\
\midrule
\texttt{(inherit A B)} & \texttt{<A --> B>} & A is a kind of B \\
\texttt{(similar A B)} & \texttt{<A <-> B>} & A resembles B \\
\texttt{(imply A B)} & \texttt{<A ==> B>} & If A then B (eternal) \\
\texttt{(predict A B)} & \texttt{<A =/> B>} & If A then B (temporal) \\
\texttt{(equiv A B)} & \texttt{<A <=> B>} & A iff B \\
\texttt{(instance A B)} & \texttt{<A |-> B>} & Instance relation \\
\bottomrule
\end{tabular}
\end{table}

\subsection{Connectors}

Connectors build compound terms. Table~\ref{tab:ds_connectors} lists all 14
connectors with their Narsese equivalents and arity constraints.

\begin{table}[h]
\centering
\caption{DriftScript connectors}
\label{tab:ds_connectors}
\begin{tabular}{llcl}
\toprule
\textbf{DriftScript} & \textbf{Narsese} & \textbf{Arity} & \textbf{Notes} \\
\midrule
\texttt{seq} & \texttt{\&/} & 2--3 & Sequential conjunction \\
\texttt{and} & \texttt{\&\&} & 2 & Conjunction \\
\texttt{or} & \texttt{||} & 2 & Disjunction \\
\texttt{not} & \texttt{-{}-} & 1 & Negation \\
\texttt{product} & \texttt{*} & 1+ & Product \\
\texttt{ext-set} & \texttt{\{\ldots\}} & 1+ & Ext.\ set \\
\texttt{int-set} & \texttt{[\ldots]} & 1+ & Int.\ set \\
\texttt{ext-inter} & \texttt{\&} & 2 & Ext.\ intersection \\
\texttt{int-inter} & \texttt{|} & 2 & Int.\ intersection \\
\texttt{ext-diff} & \texttt{-} & 2 & Ext.\ difference \\
\texttt{int-diff} & \texttt{\~{}} & 2 & Int.\ difference \\
\texttt{ext-image1/2} & \texttt{/1, /2} & 2 & Ext.\ image \\
\texttt{int-image1/2} & \texttt{\textbackslash 1, \textbackslash 2} & 2 & Int.\ image \\
\bottomrule
\end{tabular}
\end{table}

\subsection{The \texttt{call} Shorthand}

Operations are invoked through \texttt{call}, which is syntactic sugar for the
standard Narsese operation encoding:

\begin{lstlisting}[language=DriftScript]
(call ^press)                        ; bare: ^press
(call ^goto (ext-set "SELF") "park") ; <(*, {SELF}, park) --> ^goto>
\end{lstlisting}

\subsection{Variables}

DriftScript supports three variable types, roughly corresponding to NAL-6
quantifiers (see \citet{wang2013non}, Ch.~7 for the precise semantics of NARS
variables):

\begin{table}[h]
\centering
\caption{DriftScript variable types}
\label{tab:vars}
\begin{tabular}{llll}
\toprule
\textbf{Prefix} & \textbf{Type} & \textbf{Narsese} & \textbf{NAL role} \\
\midrule
\texttt{\$} & Independent & \texttt{\$1}, \texttt{\$2} & Universal-like \\
\texttt{\#} & Dependent & \texttt{\#1}, \texttt{\#2} & Existential-like \\
\texttt{?} & Query & \texttt{?1}, \texttt{?2} & ``What fills this?'' \\
\bottomrule
\end{tabular}
\end{table}

Descriptive names (\texttt{\$animal}, \texttt{?what}) are automatically mapped
to numbered Narsese variables. The mapping resets per top-level form. The
compiler pre-scans for explicitly numbered variables and assigns named
variables to the next available slot, preventing collisions.

\subsection{Quoting Rules}

All concept names must be double-quoted. Keywords, operations, and variables
must \emph{not} be quoted. This eliminates ambiguity between language structure
and user data. Strings support \texttt{\textbackslash"} and
\texttt{\textbackslash\textbackslash} escapes.

\subsection{Meta Commands}

Meta commands compile to engine control directives:

\begin{lstlisting}[language=DriftScript]
(cycles 10)                  ; run 10 inference cycles
(def-op ^press)              ; register an operation
(reset)                      ; clear all memory
(config volume 0)            ; set configuration parameter
(concurrent)                 ; mark next input as simultaneous
\end{lstlisting}

\section{Compiler Architecture}
\label{sec:compiler}

\subsection{Overview}

The DriftScript compiler is a pure text transformer with no dependency on the
DriftNARS engine. It is behaviour-preserving with respect to emitted Narsese and engine
execution: it emits standard Narsese forms accepted by DriftNARS/ONA-compatible
engines without modifying inference behaviour. This is validated empirically
through equivalence testing (Section~\ref{sec:eval}) rather than proven
formally.

The compiler is 1{,}941 lines of C99 with zero dependencies, following a
four-stage pipeline where each stage produces a separate representation:

\begin{center}
Source $\rightarrow$ \fbox{Tokeniser} $\rightarrow$ Tokens $\rightarrow$
\fbox{Parser} $\rightarrow$ AST $\rightarrow$ \fbox{Compiler} $\rightarrow$
\fbox{Emitter} $\rightarrow$ Output
\end{center}

\subsection{Tokeniser}

Converts source into up to 1{,}024 tokens of five types:
\texttt{TOK\_LPAREN}/\texttt{TOK\_RPAREN}, \texttt{TOK\_KEYWORD}
(colon-prefixed), \texttt{TOK\_STRING} (quoted with escape support), and
\texttt{TOK\_SYMBOL}. Line/column tracking enables precise diagnostics.

\subsection{Parser}

A recursive-descent parser builds an AST with two node types:
\texttt{NODE\_ATOM} (leaf with value and \texttt{quoted} flag) and
\texttt{NODE\_LIST} (up to 16 children). Nodes are allocated from a fixed
pool of 2{,}048 entries. These static bounds are generous for incremental
agent scripting but would not accommodate very large batch inputs; large-scale
batch compilation would require dynamic allocation.

\subsection{Compiler}

Each top-level form yields a \texttt{DS\_CompileResult} tagged with a result
kind:

\begin{lstlisting}[language=C]
typedef enum {
    DS_RES_NARSESE,        // Narsese sentence
    DS_RES_SHELL_COMMAND,  // Engine directive
    DS_RES_CYCLES,         // Cycle count
    DS_RES_DEF_OP          // Operation registration
} DS_ResultKind;
\end{lstlisting}

The result kind allows host environments to route results without parsing
output strings. Compilation dispatches on the head symbol: sentence forms
recursively compile terms, copula/connector handlers validate arity, and
variable renaming pre-scans for reserved numbers. Validation is pervasive:
arity, truth ranges, tense legality, quoting rules, and config keys are all
checked with source-location diagnostics.

\subsection{Emitter}

In standalone mode, results are written to stdout. In library mode
(\texttt{DS\_LIBRARY}), results are returned through:

\begin{lstlisting}[language=C]
int DS_CompileSource(const char *source,
    DS_CompileResult *results, int max_results);
\end{lstlisting}

\section{Integration Architecture}
\label{sec:integration}

DriftScript is a standalone compiler. The integration mechanisms described
here are part of the DriftNARS engine \citep{brady2024driftnars}, not
DriftScript itself. They are relevant because they enable DriftScript programs
to participate in agent loops.

\subsection{Callback Types}

DriftNARS provides four callback types with flat C primitives for FFI
compatibility:

\textbf{Event Handler} --- fires for input/derived/revised events.
\textbf{Answer Handler} --- fires when a question is answered.
\textbf{Decision Handler} --- fires on decisions above threshold, exposing
the implication, precondition, and expectation.
\textbf{Execution Handler} --- fires when an operation executes:

\begin{lstlisting}[language=C]
typedef void (*NAR_ExecutionHandler)(void *userdata,
    const char *op, const char *args);
\end{lstlisting}

\subsection{HTTP Operation Registry}

The DriftNARS HTTP server maps operation names to callback URLs. On execution,
the server POSTs a JSON payload:

\begin{lstlisting}[language={}]
{"op":"^press", "args":"({SELF} * switch)",
 "frequency":1.0, "confidence":1.0,
 "timestamp_ms":1711584000000}
\end{lstlisting}

External systems participate in the agent loop via standard HTTP.

\subsection{Python Bindings}

The Python wrapper (246 lines, ctypes) provides \texttt{on\_event},
\texttt{on\_answer}, \texttt{on\_decision}, \texttt{on\_execution} callbacks
and \texttt{add\_driftscript()} for compiling and dispatching DriftScript.

\subsection{The Sense-Reason-Act Loop}

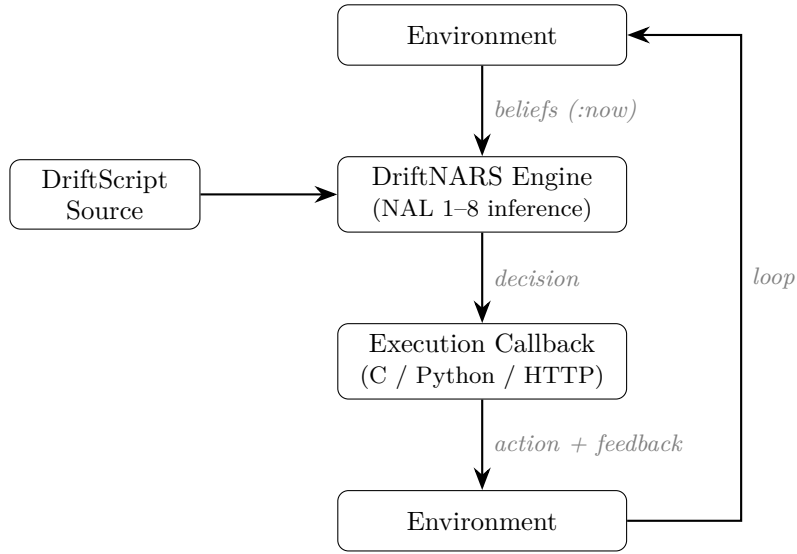
\begin{figure}[h]
\centering
\begin{tikzpicture}[
    node distance=1.2cm,
    box/.style={rectangle, draw, rounded corners, minimum width=3.8cm,
                minimum height=0.8cm, align=center, font=\small},
    arrow/.style={-{Stealth[length=3mm]}, thick},
    label/.style={font=\footnotesize\itshape, text=gray}
]
\node[box] (env1) {Environment};
\node[box, below=of env1] (engine) {DriftNARS Engine\\{\footnotesize(NAL 1--8 inference)}};
\node[box, below=of engine] (callback) {Execution Callback\\{\footnotesize(C / Python / HTTP)}};
\node[box, below=of callback] (env2) {Environment};
\node[box, left=1.8cm of engine, minimum width=2.5cm] (ds) {DriftScript\\Source};

\draw[arrow] (env1) -- node[right, label] {beliefs (:now)} (engine);
\draw[arrow] (ds) -- (engine);
\draw[arrow] (engine) -- node[right, label] {decision} (callback);
\draw[arrow] (callback) -- node[right, label] {action + feedback} (env2);
\draw[arrow] (env2.east) -- ++(1.5,0) |- node[right, label, pos=0.25] {loop} (env1.east);
\end{tikzpicture}
\caption{Integration of DriftScript with DriftNARS in a sense--reason--act loop.}
\label{fig:loop}
\end{figure}

\section{Evaluation}
\label{sec:eval}

\subsection{Compiler Correctness}

The compiler includes a test suite of \textbf{106 test cases} across 13
categories (Table~\ref{tab:tests}). All 106 pass. Tests include unit tests
for each compiler stage and integration tests comparing emitted Narsese
against expected canonical outputs.

\begin{table}[h]
\centering
\caption{Compiler test suite summary}
\label{tab:tests}
\begin{tabular}{lrl}
\toprule
\textbf{Category} & \textbf{Tests} & \textbf{Coverage} \\
\midrule
Tokeniser & 14 & Parens, keywords, strings, escapes, comments \\
Parser & 6 & Atoms, nested lists, errors \\
Copulas & 8 & All 6 copulas + arity errors \\
Connectors & 17 & All 14 connectors + arity errors \\
Call shorthand & 4 & Bare ops, with args, errors \\
Sentences & 12 & All forms, tenses, truth, dt \\
Variables & 6 & Named, numbered, collision avoidance \\
Meta commands & 7 & cycles, reset, def-op, config \\
Nested compounds & 5 & Deep nesting, mixed forms \\
Multi-statement & 2 & Multiple forms, scope isolation \\
Error detection & 4 & Bare atoms, unknown forms \\
Quoting enforcement & 7 & All violation categories \\
Validation & 10 & Range checks, illegal combinations \\
\midrule
\textbf{Total} & \textbf{106} & \\
\bottomrule
\end{tabular}
\end{table}

\subsection{Equivalence Testing}

To verify that compiled DriftScript produces identical engine behaviour to
hand-written Narsese, we ran matched inputs through the DriftNARS engine and
compared outputs.

\textbf{Deduction chain.} Both paths produce:
\texttt{Answer: <robin --> animal>. Truth: frequency=1.000000,
confidence=0.810000} with identical stamps and creation times.

\textbf{Temporal rule with goal-directed execution.} Both paths produce:
\texttt{decision expectation=0.791600} with identical implication truth values,
precondition stamps, and \texttt{\^{}press executed} output.

In both cases engine output is \textbf{byte-identical}. We tested 12
representative programs spanning deduction, temporal inference, variable use,
and operation invocation; all produced byte-identical outputs. The compiler is
transparent to the reasoning engine.

\subsection{NAL Coverage Analysis}

Table~\ref{tab:coverage} maps NAL levels to DriftScript constructs.

\begin{table}[h]
\centering
\caption{NAL coverage. All entries marked ``Covered'' have compiler test
cases and produce correct Narsese output.}
\label{tab:coverage}
\begin{tabular}{llll}
\toprule
\textbf{NAL} & \textbf{Construct} & \textbf{DriftScript} & \textbf{Notes} \\
\midrule
1 & Inheritance & \texttt{inherit} & \\
2 & Similarity & \texttt{similar} & \\
2 & Instance & \texttt{instance} & Sugar for \texttt{|->} \\
3 & Set operations & \texttt{ext-set}, \texttt{int-set} & \\
3 & Intersection/diff & \texttt{ext-inter}, etc. & 4 operators \\
4 & Product & \texttt{product} & N-ary \\
4 & Image & \texttt{ext-image1/2}, etc. & 4 operators \\
5 & Logical connectives & \texttt{and}, \texttt{or}, \texttt{not} & \\
5 & Implication & \texttt{imply} & Eternal \\
5 & Equivalence & \texttt{equiv} & \\
6 & Variables & \texttt{\$}, \texttt{\#}, \texttt{?} & Auto-numbered \\
7 & Tense & \texttt{:now}, \texttt{:past}, \texttt{:future} & \\
7 & Temporal impl. & \texttt{predict} & \\
7 & Temporal offset & \texttt{:dt} & \\
8 & Seq.\ conjunction & \texttt{seq} & Max 3 elements \\
8 & Operations & \texttt{call} & \\
8 & Goals & \texttt{goal} & Always present \\
\bottomrule
\end{tabular}
\end{table}

The covered constructs correspond to those commonly used in practical NAL
programs. Higher-order statements (implications as terms in copulas) can be
represented by nesting, but DriftScript does not provide explicit syntax for
all possible combinations. Less common or highly nested forms can be expressed
via raw Narsese, which can be freely mixed with DriftScript input.

\subsection{Structural Readability Metrics}

We compiled 15 representative programs in both DriftScript and Narsese and
measured structural properties (Table~\ref{tab:readability}).

\begin{table}[h]
\centering
\caption{Structural comparison (15 representative programs)}
\label{tab:readability}
\begin{tabular}{lrrl}
\toprule
\textbf{Metric} & \textbf{DriftScript} & \textbf{Narsese} & \textbf{Ratio} \\
\midrule
Total characters & 693 & 412 & 1.68$\times$ \\
Symbol characters & 107 & 167 & 0.64$\times$ \\
Distinct symbols & 8 & 20 & 0.40$\times$ \\
Alphabetic chars & 452 & 186 & 2.43$\times$ \\
Alpha / total & 0.64 & 0.44 & --- \\
\bottomrule
\end{tabular}
\end{table}

DriftScript uses \textbf{36\% fewer symbolic characters} and \textbf{60\%
fewer distinct symbol types}. The 8 symbols in DriftScript (\texttt{\$ ( ) -
: ? \^{} \_}) are a strict subset of the 20 in Narsese. The alphabetic ratio
rises from 0.44 to 0.64, indicating more readable text relative to
punctuation.

These are structural metrics that quantify syntactic differences, not
cognitive load. They support the qualitative argument that DriftScript
replaces symbolic density with alphabetic keywords.

\subsection{Compilation Performance}

The compiler processes 300 forms in \textbf{3\,ms} on an Apple M-series
processor (single-threaded). In library mode, compilation is effectively instantaneous relative
to inference cycles. The compiler allocates no dynamic memory.

\section{Worked Examples}
\label{sec:examples}

\subsection{Deductive Reasoning}

\begin{lstlisting}[language=DriftScript]
(believe (inherit "robin" "bird"))
(believe (inherit "bird" "animal"))
(cycles 5)
(ask (inherit "robin" "animal"))
(cycles 5)
\end{lstlisting}

Engine output:
\begin{lstlisting}[language={},basicstyle=\footnotesize\ttfamily]
Answer: <robin --> animal>. creationTime=2 Stamp=[2,1]
  Truth: frequency=1.000000, confidence=0.810000
\end{lstlisting}

Confidence decreases from 0.9 to 0.81 because the conclusion rests on an
inference chain rather than direct evidence.

\subsection{Temporal Learning and Goal-Directed Action}

\begin{lstlisting}[language=DriftScript]
(def-op ^grab)
(config volume 0)

;; Round 1: observe
(believe "see_food" :now) (cycles 5)
(believe (call ^grab) :now) (cycles 5)
(believe "have_food" :now) (cycles 5)

;; Round 2: reinforce
(believe "see_food" :now) (cycles 5)
(believe (call ^grab) :now) (cycles 5)
(believe "have_food" :now) (cycles 5)

;; Apply learned rule
(believe "see_food" :now)
(goal "have_food")
(cycles 10)
;; Engine executes ^grab via learned rule
\end{lstlisting}

No explicit rules were provided. The engine observed the pattern twice,
formed \texttt{<(see\_food \&/ \^{}grab) =/> have\_food>}, and executed
\texttt{\^{}grab} when the precondition was true and the goal active.

\subsection{Multi-Step Planning with Callbacks}

\begin{lstlisting}[language=Python]
from driftnars import DriftNARS

with DriftNARS() as nar:
    state = {"location": "home", "has_key": False}

    def handle_execution(op, args):
        if op == "^pickup":
            state["has_key"] = True
            nar.add_narsese("have_key. :|:")
        elif op == "^unlock" and state["has_key"]:
            state["location"] = "inside"
            nar.add_narsese("inside. :|:")

    nar.on_execution(handle_execution)

    nar.add_driftscript("""
        (def-op ^pickup)
        (def-op ^unlock)
        (believe (predict (seq "see_key" (call ^pickup))
                          "have_key"))
        (believe (predict (seq "have_key" (call ^unlock))
                          "inside"))
        (believe "see_key" :now)
        (goal "inside")
        (cycles 20)
    """)
\end{lstlisting}

The engine finds that \texttt{\^{}unlock} requires \texttt{have\_key},
executes \texttt{\^{}pickup} to obtain it, receives state feedback via the
callback, and then executes \texttt{\^{}unlock}. Whether multi-step
decomposition succeeds depends on cycle budget, concept priority, and
decision threshold; it is not guaranteed for arbitrary chain lengths.

\subsection{HTTP-Based Agent}

\begin{lstlisting}[language={}]
curl -X POST http://localhost:8080/ops/register \
  -d '{"op":"^water",
       "callback_url":"http://localhost:3000/water"}'

curl -X POST http://localhost:8080/driftscript \
  -d '(believe (predict (seq "soil_dry" (call ^water))
                        "soil_moist"))
      (config decisionthreshold 0.5)'

curl -X POST http://localhost:8080/narsese \
  -d 'soil_dry. :|:'
curl -X POST http://localhost:8080/narsese \
  -d 'soil_moist! :|:'
curl -X POST http://localhost:8080/narsese -d '10'
\end{lstlisting}

The entire interaction uses standard HTTP and JSON.

\section{Related Work}
\label{sec:related}

\subsection{DSLs for Logic and Reasoning}

Prolog \citep{clocksin2003programming} established logic programming with
Horn clauses. Datalog \citep{ceri1989datalog} restricts it for termination
guarantees. Answer Set Programming \citep{gebser2012answer} provides
declarative combinatorial search. These operate under the Closed World
Assumption without uncertainty or temporal reasoning.

Probabilistic languages such as ProbLog \citep{deraedt2007problog} and Church
\citep{goodman2008church} add probabilistic inference but do not address
NARS-specific concerns of resource-bounded reasoning and temporal sequencing.

DriftScript differs in targeting a specific formal language (Narsese) as
output---closer to a syntax frontend or transpiler than a standalone logic
language.

\subsection{NARS Implementations}

OpenNARS \citep{wang1995non} is the reference Java implementation. ONA
\citep{hammer2020opennars} reimplemented the core in C. Neither provides an
alternative surface syntax. We are not aware of a published compiled DSL
that covers this range of Narsese constructs.

\subsection{Agent Programming Languages}

AgentSpeak(L) \citep{rao1996agentspeak} and Jason \citep{bordini2007jason}
provide BDI agent constructs on classical logic. DriftScript occupies a
similar niche for NARS: a readable authoring surface for agent programs, with
the reasoning engine handling inference under uncertainty.

\subsection{Why S-Expressions?}

The choice of Lisp-like syntax was driven by four practical considerations:
(1)~trivial parsing with recursive descent, (2)~uniform prefix notation for
arity checking, (3)~no ambiguity with Narsese punctuation, and (4)~a natural
path to macro support.

\section{Discussion and Future Work}
\label{sec:conclusion}

\subsection{Limitations}

DriftScript is a \emph{syntactic transformation} with no formal semantics
beyond compilation to Narsese. It does not perform type checking, semantic
analysis, or optimisation. The callback architecture is part of DriftNARS, not
DriftScript. The evaluation is structural and demonstrative rather than
comparative---we do not have user study data.

\subsection{Future Directions}

\textbf{Semantic validation} could warn on eternal temporal implications,
unregistered operations, and unreachable goals. \textbf{Inline callbacks}
could bridge the declarative/imperative boundary. A \textbf{language server}
would improve the development experience. A formal \textbf{user study} would
quantify the readability improvements we currently support only with
structural metrics.

\subsection{Conclusion}

DriftScript provides a readable, Lisp-like surface syntax for a broad subset
of Narsese covering NAL~1--8. Its compiler passes 106 tests, produces output
byte-identical to hand-written Narsese, reduces symbolic density by 36\%, and
compiles 300 forms in 3\,ms. Combined with DriftNARS's callback architecture,
it supports agent development through C, Python, and HTTP interfaces.

DriftScript does not change NARS's reasoning---it provides a more readable
authoring surface for the programs that drive it.


\bibliographystyle{plainnat}

\appendix
\section{Translation Examples}
\label{app:translations}

Table~\ref{tab:translations} shows 20 representative DriftScript forms and
their compiled Narsese output, verified by the test suite.

\begin{table*}[h]
\centering
\caption{DriftScript to Narsese translation examples}
\label{tab:translations}
\footnotesize
\begin{tabular}{rll}
\toprule
\textbf{\#} & \textbf{DriftScript} & \textbf{Narsese} \\
\midrule
1  & \texttt{(believe (inherit "robin" "bird"))} & \texttt{<robin --> bird>.} \\
2  & \texttt{(believe (similar "cat" "dog"))} & \texttt{<cat <-> dog>.} \\
3  & \texttt{(believe (imply "rain" "wet"))} & \texttt{<rain ==> wet>.} \\
4  & \texttt{(believe (predict "A" "B"))} & \texttt{<A =/> B>.} \\
5  & \texttt{(believe (equiv "A" "B"))} & \texttt{<A <=> B>.} \\
6  & \texttt{(believe (instance "Tweety" "bird"))} & \texttt{<Tweety |-> bird>.} \\
7  & \texttt{(believe "light\_on" :now)} & \texttt{light\_on. :|:} \\
8  & \texttt{(goal "light\_off")} & \texttt{light\_off! :|:} \\
9  & \texttt{(ask (inherit "robin" "animal"))} & \texttt{<robin --> animal>?} \\
10 & \texttt{(ask (inherit ?x "animal"))} & \texttt{<?1 --> animal>?} \\
11 & \texttt{(believe (predict (seq "a" "b") "c"))} & \texttt{<(a \&/ b) =/> c>.} \\
12 & \texttt{(believe (inherit (ext-set "SELF") "agent"))} & \texttt{<\{SELF\} --> agent>.} \\
13 & \texttt{(believe (inherit "x" (int-set "bright")))} & \texttt{<x --> [bright]>.} \\
14 & \texttt{(believe (inherit (product "A" "B") "rel"))} & \texttt{<(*, A, B) --> rel>.} \\
15 & \texttt{(believe (imply (inherit \$x "bird") ...)} & \texttt{<<\$1 --> bird> ==> <\$1 --> animal>>.} \\
16 & \texttt{(believe (predict (seq ... (call \^{}press)) ...))} & \texttt{<(light\_on \&/ \^{}press) =/> light\_off>.} \\
17 & \texttt{(call \^{}goto (ext-set "SELF") "park")} & \texttt{<(*, \{SELF\}, park) --> \^{}goto>} \\
18 & \texttt{(believe (and "A" "B"))} & \texttt{(A \&\& B).} \\
19 & \texttt{(believe (not "A"))} & \texttt{(-{}- A).} \\
20 & \texttt{(believe (or "A" "B"))} & \texttt{(A || B).} \\
\bottomrule
\end{tabular}
\end{table*}

\end{document}